\documentclass[%
 reprint,
 amsmath,amssymb,
 aps,prl
]{revtex4-2}

\usepackage{graphicx}
\usepackage{dcolumn}
\usepackage{bm}
\usepackage{color}

\usepackage[colorinlistoftodos]{todonotes}
\usepackage{comment}
\usepackage{standalone}
\usepackage{lipsum, babel}

\begin{document}


\title{Confined phase singularities reveal the spin-to-orbital angular momentum conversion of sound waves} 


\author{Ludovic Alha{\"{\i}}tz}
\affiliation{
Univ. Bordeaux, CNRS, Bordeaux INP, I2M, UMR 5295, F-33400, Talence, France}%
 
\author{Thomas Brunet}
\affiliation{
Univ. Bordeaux, CNRS, Bordeaux INP, I2M, UMR 5295, F-33400, Talence, France}%

\author{Christophe Aristégui}
\affiliation{
Univ. Bordeaux, CNRS, Bordeaux INP, I2M, UMR 5295, F-33400, Talence, France}%

\author{Olivier Poncelet}
\affiliation{
Univ. Bordeaux, CNRS, Bordeaux INP, I2M, UMR 5295, F-33400, Talence, France}%

\author{Diego Baresch}
\email{diego.baresch@u-bordeaux.fr}
\affiliation{
Univ. Bordeaux, CNRS, Bordeaux INP, I2M, UMR 5295, F-33400, Talence, France}%

\email{diego.baresch@u-bordeaux.fr and olivier.poncelet@u-bordeaux.fr}


\begin{abstract}
We identify an acoustic process in which the conversion of angular momentum between its spin and orbital form takes place. The interaction between an evanescent wave propagating at the interface of two immiscible fluids and an isolated droplet is considered. The elliptical motion of the fluid supporting the incident wave is associated with a simple state of spin angular momentum, a quantity recently introduced for acoustic waves in the literature. We experimentally observe that this field predominantly forces a directional wave transport circling the droplet's interior, revealing the existence of confined phase singularities. The circulation of the phase, around a singular point, is characteristic of angular momentum in its orbital form, thereby demonstrating the conversion mechanism. The numerical and experimental observations presented in this work have implications for the fundamental understanding of the angular momentum of acoustic waves, and for applications such as particle manipulation with radiation forces or torques, acoustic sensing and imaging.     
\end{abstract}

\maketitle

It is widely accepted that the longitudinal nature of sound waves cannot provide for a spin angular momentum (AM) density in the supporting liquid or gas. Indeed, in the absence of an efficient transverse restoring mechanism, the motion of a lossless fluid carrying sound waves is often thought to remain exclusively aligned with the propagation direction, and therefore lacks the kind of rotational degrees of freedom that leads to spin AM. While this is easily verified for the ideal case of homogeneous plane waves, the actual fluid motion in \textit{e.g.} non-paraxial beams, interfering waves or surface waves is more complex, and the existence of a finite spin AM density has recently been suggested \citep{Long2018,Shi2019,Bliokh2019a,Bliokh2019b}.  

For light, in contrast, the separation of AM into a spin part associated with the circular transverse polarisation of the electric field \citep{Beth1936}, and  an orbital part associated with the spatial distribution of the phase \citep{Allen1992} is well established and has been confirmed by several experiments two decades ago \citep{Simpson1997, Oneil2002, Padgett2003}. Nevertheless, a recent interest in these quantities in the context of highly structured or confined optical fields has revealed other intriguing properties of the AM of light, generically termed spin-orbit interactions, provided the full vectorial character of Maxwell's equations is considered, comprising polarisation components of the electric and magnetic fields pointing in the direction of propagation, transverse components of the spin AM density  \citep{Bliokh2014,Neugebauer2015,Bliokh2015}, conversion mechanisms between the two forms of AM \citep{Zhao2007,Bliokh2011spin} and interaction rules with quantum emitters \cite{Andersen2006,Mitsch2014}.

In simple lossless fluids, longitudinal sound waves are frequently termed `pressure' or `scalar' waves, and the analysis of the vectorial properties of the velocity field are generally overlooked, with the exception of a few reported seminal studies \citep{Hayes1986, Deschamps1989}, and more recent accounts on polarisation singularities and topological properties of sound \cite{Bliokh2021polarization, Muelas2022}. In particular, this may explain why only AM in its orbital form, which arises from specific spatial phase properties of a scalar potential field, has been previously investigated in acoustic vortex beams \citep{Hefner,Thomas2003,Volke-Sepulveda2008,Anhauser2012}. Consequently, most accounts of `spinlike' states of sound, or spin-orbit interactions have been reported in artificial elastic structures \citep{He2016,Ge2021,Fan2021,Wang2021}. For transverse elastic waves in crystals (phonons) - known to share several properties with light (photons) \citep{Levine1962,Mclellan1988} - the experimental evidence of an existing spin AM is only very recent \citep{Zhang2014,Holanda2018}. 

The renewed interest in the fundamental nature of the AM density of sound waves in simple fluids, with the recent proposition of its separation between well-defined orbital and spin components, calls for the development of experiments sensitive to both forms of AM through direct wavefield measurements or the mechanical evidence of their transfer to matter \citep{Bliokh2019c,Lopes2020}. Strictly speaking, we note that acoustic waves are characterised by an angular `pseudomomentum', since the medium itself is not subject to a net transport of mass \citep{McIntyre1981}.   

Here we identify a process by which the spin AM density of a propagating longitudinal sound wave is converted into AM in its orbital form. We characterise the interaction between an isolated fluid droplet and an incident evanescent (inhomogenous) field. By exploiting minimally invasive and high resolution acoustic pressure measurements, we were able to map the spatial properties of the internal acoustic field. Our measurements reveal the emergence of wavefronts circulating around phase singularities confined in mm-sized resonant droplets. The circulation of the  wave's phase around singular points is a hallmark for orbital AM imparted to the fluid, which is found to co-exist with a non-trivial spin AM density inside the droplet. Our observations suggest spin-orbit interactions are ubiquitous when inhomogenous sound waves interact with simple targets.

\paragraph{Transverse spin density of the incident evanescent wave}

\begin{figure}
\centering \includegraphics[width=1.0\linewidth]{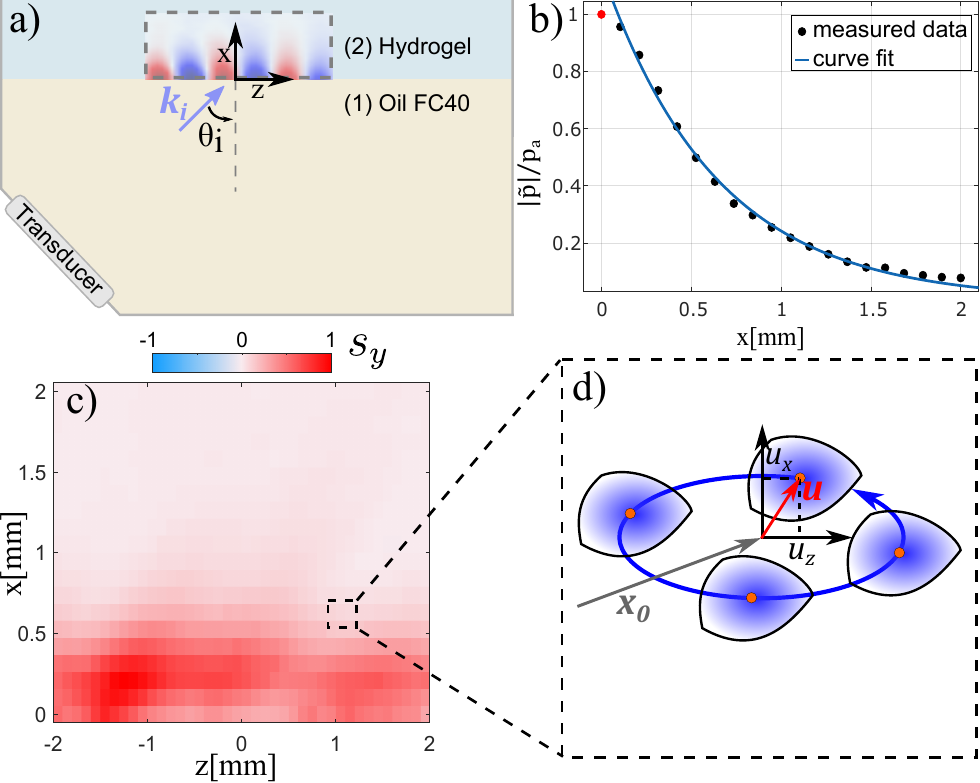}
\caption{a) A plane wave generated by a piezoelectric transducer in the `slow' medium (1) made of fluorinated oil impinges the interface with the `fast' medium (2) (hydrogel) at a supercritical reflection angle. The experimental wave generated in medium (2) is shown in the inset and propagates along $z$. b) shows the exponential decay of its amplitude along $x$. c) Transverse spin AM density, $s_y$, measured in a region where the evanescent wave propagates (normalised to unity). d) Schematic representation of the periodic elliptical motion of a fluid particle about its position at rest $\mathbf{x}_0$.}
\label{fig1}
\end{figure}

In a lossless fluid of mass density $\rho$, the spin AM density of a sound wave was recently derived as \cite{Shi2019,Bliokh2019a, Bliokh2019b}:

\begin{equation}
\mathbf{s}= \frac{\rho}{2\omega}\text{Im}(\mathbf{v}^*\times \mathbf{v})=\frac{1}{2\rho\omega^3}\text{Im}(\mathbf{\nabla}p^*\times \mathbf{\nabla}p), 
        \label{eq1}
\end{equation}
where $\mathbf{v}$ is the complex velocity field in an Eulerian description of the fluid motion, $^*$ denotes its complex conjugate, Im denotes the imaginary part,  and Euler's linearised momentum equation, $\mathrm{i}\omega \rho \mathbf{v}=\mathbf{\nabla}p$ was used for a monochromatic pressure field, $p$, varying in time, $t$ as $\mathrm{e}^{-\mathrm{i} \omega t}$ with a frequency, $f=\omega/2\pi$. Evanescent plane waves are an example of  inhomogeneous acoustic fields that give rise to a simple state of transverse spin AM \cite{Bliokh2019a}. They are plane wave solutions  of the propagation equation obtained using a complex-valued wave vector $\mathbf{k} = k_z \mathbf{e}_z + \mathrm{i}\kappa \mathbf{e}_x$. The pressure field can be written as 
\begin{equation}
    p(x,z,t)=p_a \mathrm{e}^{ -\kappa x}\mathrm{e}^{\mathrm{i}(k_z z-\omega t)},
    \label{eq2}
\end{equation}
where $x$ denotes the vertical direction (unit vector $\mathbf{e}_x$) along which the wave amplitude decays exponentially, $z$ denotes the propagation direction (unit vector $\mathbf{e}_z$), and $p_a$ is a reference acoustic pressure amplitude (see Figure \ref{fig1}). The solution is invariant in the $y$-direction. The dispersion relation $\mathbf{k}\cdot\mathbf{k}=k_z^2-\kappa^2=k^2$ relates the propagation wavevector $k_z$ to the spatial decay rate $\kappa$ via the wavenumber $k=\omega/c$, where $c$ is the speed of sound in the fluid. The local fluid displacement field is $\mathbf{u}=\mathrm{i}\omega^{-1} \mathbf{v}$.

It is important to note that the wavefield is longitudinal, \textit{i.e.} the acoustic displacement vector $\mathbf{u}$ and the wave vector $\mathbf{k}$ are collinear. However, because $\mathbf{k}$ is complex-valued, a fluid particle initially located at a position $\mathbf{x}_0$ will follow the displacement field $\mathbf{u}(\mathbf{x}_0)$ that locally describes an elliptical trajectory determined by the relative magnitude of $k_z$ and $\kappa$ \citep{Hayes1986,Deschamps1989}. It is nevertheless easy to verify that such a `polarisation' of the fluid motion does not violate its fundamental irrotational nature ($\mathbf{\nabla}\times \mathbf{u}=\mathbf{0}$), schematically illustrated by the unchanged orientation of an isolated fluid `particle' in Fig. \ref{fig1}d during its one-period  centre-of-mass trajectory. The elliptical trajectory of the fluid motion is somewhat similar to the orbits observed for gravity water waves, for which a spin density has been introduced and recently measured \cite{Jones1973, Bliokh2022}. 

To experimentally generate an evanescent acoustic wave, we designed a setup inspired by Refs.\cite{Osterhoudt2008,Plotnick2016}, that offers appropriate conditions to obtain the total internal reflection of an incident finite beam generated in medium (1) at the interface with medium (2) (Figure 1a, and Supplementary Material, Section I \footnote[1]{See Supplemental Material for additional information on the experimental setup and radiation force and torque calculations, which includes Refs.\cite{Lidon2017,Settnes2012,Karlsen2015}.}). Critical reflection conditions are obtained by using a `slow' incident medium (1) having a low speed of sound relative to the `fast' transmission medium (2). We chose for medium (1) a fluorinated oil immiscible with water (Fluorinert FC-40), of mass density ($\rho_1=1850$~kg/m$^3$) and speed of sound, $c_1=640$~m/s \citep{Tallon2017}. Medium (2) was a yield-stress gel prepared by mixing distilled water with a carbomer (Carbopol$^\text{\textregistered}$). The incident angle $\theta_i$ relative to the $x$-direction was adjusted to exceed the critical angle $\theta_c=\sin^{-1}(c_1/c_2)\sim 25$ degrees. The instantaneous pressure variation could be mapped in three-dimensions (3D) using a fibre optic hydrophone system (Precision Acoustics, UK) mounted on motorised translation stages. The fibre is $100 \, \mu$m in diameter, with a sensitive area of $10\, \mu$m that defines our resolution. Wave packets of 5 to 10 acoustic cycles were generated by the transducer (central frequency of $f=1$~MHz, with a 60\% bandwidth) and sent towards the interface. The amplitude decay in the x-direction was obtained from the Fourier transform (analysed at 1 MHz), $\tilde p$, applied to the time-dependent pressure field as shown in Fig. \ref{fig1}b (dots). The decay length was obtained from an exponential fit of the data points (plain blue curve), and estimated to be $1/\kappa\sim 0.64$~mm. The red data point was supposed to lie at the interface $x=0$ and excluded from the fit. Its value was used as the reference pressure $p_a=50 \pm 7~$~kPa.

\begin{figure*}
\centering \includegraphics[width=0.9\linewidth]{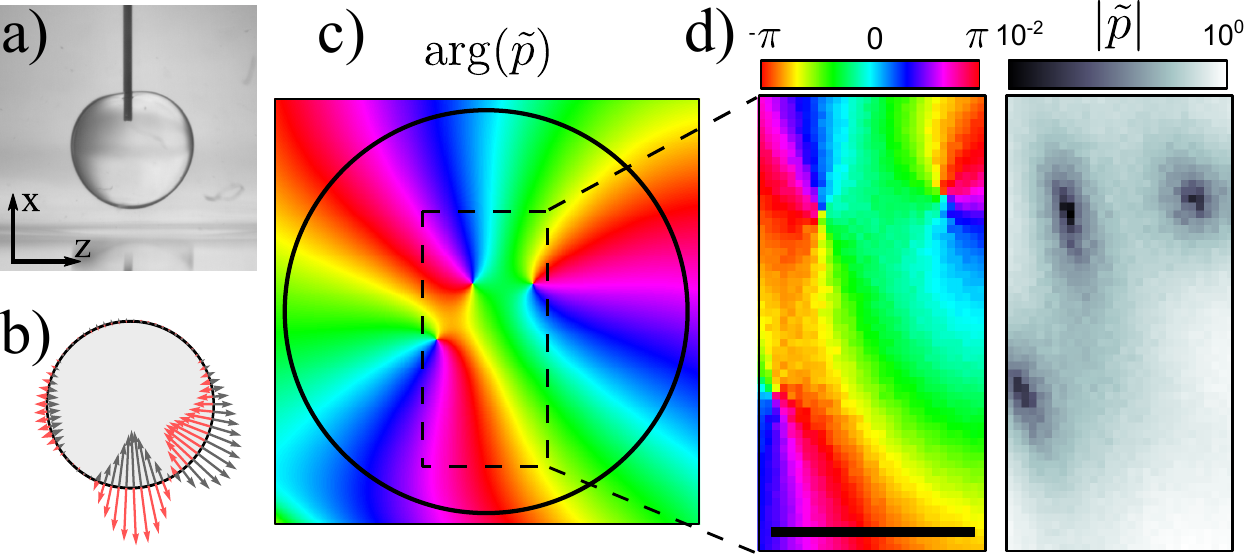}
\caption{a) Photograph showing the droplet positioned near the propagation interface and penetrated by the hydrophone. b) Normal projections of the modelled incident velocity field forcing the droplet at $t=0.2/f$ (grey) and $t=0.7/f$ (red). c) Phase field, $\mathrm{arg}(\tilde{p})$ obtained numerically inside and outside of the droplet oscillating on its hexapolar resonant mode  (for $a=1.12$~mm and $f=0.54$~MHz). d) Experimental phase field revealing three singularities confined in the droplet, and magnitude (logarithmic scale) of the experimental pressure field, $|\tilde{p}|$, dropping to zero at each singularity. Scale bar represents 500~$\mu$m.}
\label{fig2}
\end{figure*}

 A direct quantification of the spin density can be obtained from the local estimate of the acoustic velocity field from discrete pressure measurements (see Eq.\ref{eq1}). The high spatial resolution of the hydrophone was exploited to map the velocity field in the $(x,z)$ plane from pressure gradients. Figure \ref{fig1}c shows the normalised map of the transverse acoustic spin density component, $\mathrm{s}_y$, generated by the  evanescent wave. The handedness of the elliptical motion is set by the wave propagation direction (increasing $z$) and constrains the spin density, $s_y$, to be positive. Note that although Eq.(\ref{eq2}) describes a wave that does not support fluid motion in the transverse plane $(x,y)$, the finite size of the transducer can generate pressure gradients and fluid displacements along $y$. The corresponding spin densities $\mathrm{s}_x$ and $\mathrm{s}_z$ are not essential to this study and have not been characterised here.

\paragraph{Confined phase singularities in a resonant droplet}
Let us now illustrate how the incident inhomogenous wave couples with an isolated object. A single fluorinated oil droplet of radius $a$ was directly injected using a syringe in medium (2) (Fig. \ref{fig2}a). The concentration of carbomer composing the hydrogel was sufficiently high to maintain in place droplets ranging in size from a few hundreds of microns to several millimetres, but remained low enough to preserve propagation characteristics similar to those of water. This strategy has previously been proposed to trap gas bubbles \cite{Leroy2008}. A first droplet of radius $a=1.12$~mm was positioned at a distance roughly equal to the incident wave's decay length, $1/\kappa$, from the interface to ensure a strong and detectable interaction. Furthermore, the large acoustic contrast between the droplet and medium (2) favours the existence of large amplitude Mie-type scattering resonances \cite{Brunet2012} that we exploited. Calculations of the resulting acoustic field inside and outside of the droplet were obtained using a semi-analytical propagation model described elsewhere \cite{Alhaitz2021}. All numerical and experimental results are shown in the $(z,x)$ plane for $y=0$. The centre of the droplet is positioned at $x=y=z=0$ in the incident field.

Figure \ref{fig2}b shows a numerical projection of the incident velocity field normal to the droplet surface. The elliptical fluid motion characterising the inhomogenous incident wave in the outer fluid has the ability to generate non-zero normal velocity components on the lower hemisphere of the droplet. It is important to note that normal velocity components also force the upper hemisphere, but are much weaker in amplitude due to the evanescent nature of the incident field. Consequently, in clear contrast with a situation for which a droplet would be forced by a locally homogeneous (z-axisymmetric) wavefront, here the internal acoustic field is imparted with a progressive wavefront circulating anti-clockwise around the droplet centre, from the lower hemisphere where it is generated, towards the upper hemisphere. This is evidenced by the spatial properties of the wave's phase shown in Figure \ref{fig2}c and d, for the droplet forced on its hexapolar resonant mode (for $f=0.54$~MHz  or $ka\sim2.55$). The numerical phase map (Fig. \ref{fig2}c) shows a particular variation pattern  around a circular path around the droplet centre close to the drop periphery. More precisely, we observe three phase ramps from $-\pi$ to $\pi$ around three particular points where the phase is undefined. These specific locations in space, known as phase singularities, are well documented for wavefields propagating in free space \cite{Nye1947}, which include the important class of acoustical and optical vortex beams \citep{Coullet1989,Soskin1997,Yao2011,Hefner,Thomas2003}. In the present case, however, the singularities are \emph{confined} within the resonant structure, and though we note similar phase singularities should naturally emerge in optical waveguides or resonators, there are, to our knowledge, only few experiments that have succeeded in detecting such phase topologies \cite{Balistreri2000}. This is presumably due to technical difficulties in measuring electromagnetic wave properties (phase and amplitude) directly inside the structure. Here, to experimentally detect these singular regions, we manoeuvred the hydrophone through the droplet's interface while inducing a limited capillary distortion (Fig. \ref{fig2}a). The region we managed to map inside the immobilised droplet was approximately 500 $\mu$m wide and 1.2 mm high. The phase and magnitude of the field were retrieved from the instantaneous pressure measurements (Fig. \ref{fig2}d), and confirm the existence of the confined phase singularities. 

\begin{figure}
\centering \includegraphics[width=1.0\linewidth]{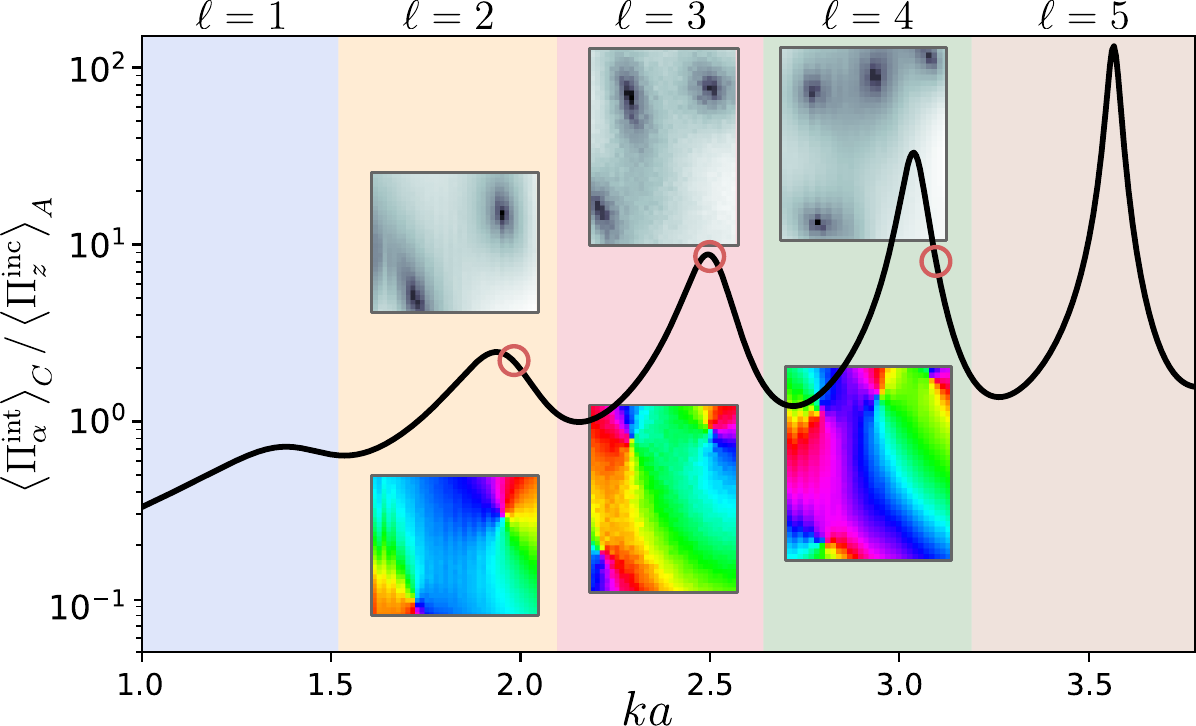}
\caption{Curve: Modelled average angular energy flux of the internal wave, $\left<\Pi^\mathrm{int}_{\alpha}\right>_C$ normalised by the average incident energy flux, $\left<\Pi^\mathrm{inc}_{z}\right>_A$. Coloured regions in $ka$-space correspond to integer values of the total topological charge, $\ell$, calculated for the numerical internal field. Insets: experimental phase and magnitude maps of the internal pressure field obtained for $ka$ values equal to 1.9, 2.55 and 3.18,  and indicated by red circles on the resonance curve. The pixel size is approximately 30~$\mu$m.}
\label{fig3}
\end{figure}

\begin{figure*}
\centering \includegraphics[width=0.95\linewidth]{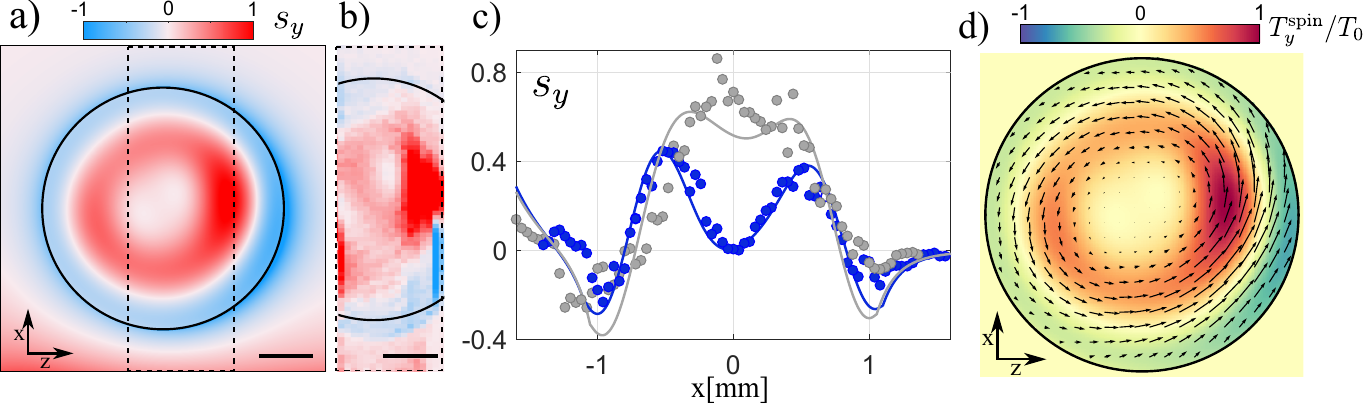}
\caption{ a) Numerical calculation of the spin AM density, $s_y$, obtained for the internal wave shown in Fig. \ref{fig2}. The data is normalised to unity and the colormap was saturated. b) Experimental measurement of the spin AM in the region delimited in a). Scale bars represent 500 $\mu$m. c) Comparison of experimental and numerical data along two distinct lines traversing the droplet vertically. d) Acoustic radiation torque, $T^\mathrm{spin}_y$, induced on an absorptive probe-like particle obtained for $a_p=$1 $\mu$m and $\bar{\rho}=2+0.5\mathrm{i}$, and normalised by the maximum computed torque, $T_0$. The arrows represent the magnitude of the tangential scattering radiation force exerted on the same particle.}   
\label{fig4}
\end{figure*}

To further explore the relationship between droplet size, frequency and phase topology, we investigated other properties of the internal wave. The energy flux, or Poynting vector, is  defined as $\mathbf{\Pi}=p\mathbf{v}$, where $p$ and $\mathbf{v}$ can either refer to the pressure and velocity fields of the incident or internal wave fields. Figure \ref{fig3} shows a calculation of the angular component of the energy flux carried by the internal field, $\left<\Pi^\mathrm{int}_{\alpha}\right>_C$, which has been normalised by the energy flux carried by the incident wave along the propagation direction, $\left<\Pi^\mathrm{inc}_z\right>_A$. The latter was averaged over the cross-section $A=\pi a^2$ orientated perpendicular to $z$, whereas the internal angular energy flux was averaged over the circular path $C=2\pi a$ with an angle $\alpha$ defined between the horizontal and vertical axes passing through the droplet centre, and varying from 0 to $2\pi$. Doing so, the circulating energy flux reveals clear amplification peaks coinciding with the resonant modes of the droplet, for which the forcing of the droplet is efficiently transferred to circulating internal modes. Remarkably, each peak is located in regions of $ka$-space characterised by an integer value of the total topological charge defined as $\ell=1/2\pi \oint_C \mathrm{d} \chi$, where $\chi=\mathrm{arg}(\tilde{p})$ \cite{dennis2009singular}. The insets in Fig. \ref{fig3} show experimental phase and magnitude maps that we were able to measure exploiting the position stability of two other droplets and the bandwidth of the transducer (See Supplemental Material \footnote[1]{}. Our summarised measurements for the approximate values $ka=1.9,  2.55$ and 3.18  unambiguously show that the integer total topological charge $\ell$ originates from  $\ell$ co-existing, but not co-localised, phase singularities of topological charge $1$, and coincides with the resonance order of the droplet, \textit{i.e.}, quadrupolar ($\ell=2$), hexapolar $(\ell=3)$ and octopolar ($\ell=4$). These locations are regions where the head pulses of the circulating wavefront interfere destructively with pulses that were delayed within the incident wave-train and interact with the droplet latter. The good timing for these arrivals underpin the appearance of resonance peaks in $\left<\Pi^\mathrm{int}_{\alpha}\right>_C$.

\paragraph{Spin-to-orbital angular momentum conversion}
We now analyse the AM density that is imparted to the fluid by the internal wave described in the previous section for the hexapolar mode (topological charge $\ell=3$). The spin AM density can again be obtained from the highly resolved pressure measurements performed inside the droplet. Remarkably, the spin AM topology inside the droplet is non-trivial (Fig. \ref{fig4}a-c). Its sign changes as the radial distance from the droplet centre is increased. This means that, in contrast with the elliptical polarisation of the fluid motion characterising the incident wave (Fig. \ref{fig1}c-d), the handedness of the polarisation inside the droplet can switch from clockwise to anti-clockwise in regions of negative and positive spin AM densities respectively. These oscillations and changes in sign can be better appreciated in Fig. \ref{fig4}c, where the spin AM density is plotted along two arbitrarily chosen vertical lines traversing the droplet.  

Importantly, it is also expected that the circulation of the wave's phase around singularities as described in Fig. \ref{fig2}c-e  will give rise to a non-zero orbital AM density, in analogy with optical or acoustical vortices \cite{Allen1992,Thomas2003}.  According to Ref.\cite{Bliokh2019b},  the orbital AM density can be computed as $\mathbf{l}=\mathbf{r} \times \mathbf{p}$, where $\mathbf{p} = \frac{\rho}{2\omega}\text{Im}[\mathbf{v}^*\cdot (\mathbf{\nabla})\mathbf{v}]$ is the canonical momentum density of the internal wave, and $\mathbf{r}$ the position vector whose origin is the droplet centre. To quantify the AM conversion between the spin-carrying  incident wave and the internal circulating hexapolar mode, we computed the integral values of the spin and orbital AM densities generated inside the droplet volume from the numerical data. The OAM to spin ratio of their integral values is approximately equal to 92\%, meaning most of  the AM exists in its orbital form, confirming an efficient transfer of momentum to the internal circulating mode. We additionally found that this ratio  remains similar for the other resonant modes.

Finally, we discuss the mechanical effects that would be induced by the two distinctive forms of AM on an absorptive probe particle  placed inside the droplet. For a small subwavelength particle (radius $a_p \ll \lambda =2\pi/k_z$),  a radiation torque will be exerted and has been related in Ref.\citep{Bliokh2019c} to the local value of the transverse spin density by the simple equation, $T^\mathrm{spin}_y=\omega \text{Im}(\alpha_d) s_y$ , where $\alpha_d$ is the dipolar polarizability coefficient of the particle (See also Supplementary Materiel, Sec II \footnote[1]{} for details on the torque calculation). As seen in Fig. \ref{fig4}d, the torque's direction is, as expected, directly related to the sign of the spin AM, and the particle would rotate about its own axis clockwise or anti-clockwise, depending on its position within the droplet. Simultaneously, the OAM is associated with a scattering radiation force, $\mathbf{F}^\mathrm{scat}$ (Supp. Mat., Sec II \footnote[1]{}), that accelerates the particle (note that a similar mechanical effect has been described for particles in acoustic vortex beams \citep{BareschJASA}). Its tangential component is represented by arrows in Fig. \ref{fig4}d, and suggests the particle will be set on an anti-clockwise orbital trajectory around the droplet centre. This kinematics can alternatively be understood as the result of an applied  torque, $T^\mathrm{scat}_y$,  about the $y$-axis passing through the drop centre, and resulting from the cross product of the particle position vector and tangential component of the scattering force, $T^\mathrm{scat}_y=zF^\mathrm{scat}_x - x F^\mathrm{scat}_z$. We find that at the position maximising $T^\mathrm{spin}_y$, the ratio between both torques $T^\mathrm{scat}_y/T^\mathrm{spin}_y$ approximates 91\%, consistent with the dominant fraction of OAM in the droplet.

\paragraph{Conclusion and perspectives}
Our results suggest that the conversion of AM between its spin and orbital forms may be very common when an inhomogenous wavefront interacts with targets of simple geometrical forms, impinging a clear acoustic signature to the internal field: the presence of phase singularities. This work also shows that complex spin ``textures" can be accurately measured in space, and therefore prepares for further investigations on the central role the spin AM density can have in inducing mechanical rotation, a role that has to date been overlooked in previous studies \citep{Volke-Sepulveda2008,Skeldon2008,Anhauser2012,Baresch2018}. Extending the field analysis to objects having other geometries, or with passive \cite{Wunenburger2015} or active \cite{Nassar2020} ``chiral" scattering properties could also reveal other routes to the conversion of angular momentum.

Several applications including acoustic imaging and sensing, particle manipulation \citep{Baresch2016,Baresch2018,Marzo2018,Bruus2011}, or developing devices for `chiral' wave guiding  \citep{He2016,Ge2021,Fan2021,Wang2021,Lanoy2020} are expected to benefit from a better understanding of the AM properties of sound waves.

\clearpage

\onecolumngrid

\section{Supplementary Material}

\title{Confined Phase Singularities Reveal the Spin-to-Orbital Angular Momentum Conversion of Sound Waves} 


\author{Ludovic Alha{\"{\i}}tz}
\affiliation{
Université Bordeaux, CNRS, Bordeaux INP, ENSAM, INRAE, UMR 5295 I2M, F-33405 Talence, France}%
 
\author{Thomas Brunet}
\affiliation{
Université Bordeaux, CNRS, Bordeaux INP, ENSAM, INRAE, UMR 5295 I2M, F-33405 Talence, France}%

\author{Christophe Aristegui}
\affiliation{
Université Bordeaux, CNRS, Bordeaux INP, ENSAM, INRAE, UMR 5295 I2M, F-33405 Talence, France}%

\author{Olivier Poncelet}
\affiliation{
Université Bordeaux, CNRS, Bordeaux INP, ENSAM, INRAE, UMR 5295 I2M, F-33405 Talence, France}%

\author{Diego Baresch}
\affiliation{
Université Bordeaux, CNRS, Bordeaux INP, ENSAM, INRAE, UMR 5295 I2M, F-33405 Talence, France}%

\email{diego.baresch@u-bordeaux.fr and olivier.poncelet@u-bordeaux.fr}


\maketitle


\section{Experimental set-up, fluids and droplets}

The experimental setup inspired by Refs.\cite{Osterhoudt2008,Plotnick2016} offers appropriate conditions to obtain the total internal reflection of an incident finite beam generated in medium (1) at the interface with medium (2) (see Fig. 1(a) in the main text). Critical reflection conditions are obtained by using a `slow' incident medium (1) having a low speed of sound relative to the `fast' transmission medium (2).  We chose for medium (1) a fluorinated oil immiscible with water (Fluorinert FC-40), of mass density $\rho_1=1850$~kg/m$^3$ and speed of sound, $c_1=640$~m/s \citep{Tallon2017}. Medium (2) was a yield-stress gel prepared by mixing distilled water with a carbomer (Carbopol$^\text{\textregistered}$). The low mass concentration of carbomer that was used (a 1\% mass concentration gel was diluted approximately in 5 times its volume in water) ensures that the speed of sound in the gel was only slightly higher than the speed in water. Therefore, the speed and density of the gel was assumed unchanged compared to water in our calculations ($\rho_2=1000$~kg/m$^3$ and $c_2=1490$~m/s at room temperature) \citep{Lidon2017}. Caution was taken to remove any residual air bubbles from the preparation. The two fluids constituting the propagation media were poured in a custom-made water tank. The large dimensions of the tank ensured the existence of a wide spatial domain where the evanescent wave remained weakly affected by diffraction effects and reflections against boundaries. The low carbomer concentration also allowed to ensure a good manoeuvrability of the flexible fibre optic hydrophone while providing a good position stability of millimeter-sized oil drops that where directly injected in medium (2)  with a  syringe having a flat-end 500 $\mu$m diameter tip. Nevertheless, the range of experimentally accessible $ka$ values is determined by the bandwidth of the transducer (60\%) and the droplet size. The droplet position and interface have to remain stable while the hydrophone  is introduced. This operation was hard for small droplets ($a<600$ $\mu$m) due to capillary resistance, while large drops ($a>1200$ $\mu$m) were hard to immobilise in the yield stress gel.

\section{Acoustic radiation force and torque}

Following the notations in Ref.\cite{Bliokh2019c}, the spin induced radiation torque, $\mathbf{T}^\mathrm{spin}$, and the scattering radiation force, $\mathbf{F}^\mathrm{scat}$, on a small sub-wavelength particle ($a_p \ll \lambda  $) read:
\begin{equation}
\begin{split}
\mathbf{T}^\mathrm{spin} & =  \text{Im}(\alpha_d) \mathbf{s}\\
\mathbf{F}^\mathrm{scat} & =  2\omega \left[ \text{Im}(\alpha_m) \mathbf{P}^{(p)} + \text{Im}(\alpha_d)  \mathbf{P}^{(\mathrm{v})}\right]
\end{split}
\end{equation}
where $ \mathbf{P}^{(p)}=1/(4\omega)\text{Im}(\beta p^* \mathrm{\nabla} p)$ and   $ \mathbf{P}^{(\mathrm{v})}=1/(4\omega)\text{Im}(\rho \mathrm{v}^* \cdot (\mathrm{\nabla}) \mathrm{v})$ are pressure and velocity contributions to the generalized canonical momentum, $\mathbf{P}= \mathbf{P}^{p}+\mathbf{P}^{(\mathrm{v})}$. The monopolar and dipolar polarizabilities  read respectively  $\alpha_m= 4\pi a_p^3/3(\bar{\beta}-1)$ and $\alpha_d= 4\pi a_p^3(\bar{\rho}-1)/(2\bar{\rho}+1)$ for the probe particle, for which we conserved the theoretical complex-valued compressibility $\beta_p=\beta \bar{\beta}=\beta (3+0.7\mathrm{i})$ and mass density, $\rho_p=\rho \bar{\rho} = \rho(2+0.5\mathrm{i})$ selected in Ref.\cite{Bliokh2019c}, where $\rho$ is the suspending medium density and $\beta=\rho c^2$ its compressibility. The acoustic polarizabilties are similar to those introduced in Ref.\cite{Baresch2016}, and the physical origin of their imaginary contribution can be attributed to the particle's radiation 'lag' or 'radiation friction' that is very rapidly amplified by dissipation processes  in actual thermo-viscous fluids \cite{Settnes2012,Karlsen2015}.

The  effect of the tangential scattering force on the probe particle can alternatively be described by an applied  torque, $T^\mathrm{scat}_y$,  about the $y$ axis passing through the drop centre, and resulting from the cross product of the particle position vector and tangential component of the scattering force, $T^\mathrm{scat}_y=zF^\mathrm{scat}_x - x F^\mathrm{scat}_z$.

\bibliography{bib_diego_ac}

\end{document}